\documentstyle[12pt]{article}

\newlength{\pubnumber} 

\catcode`\@=11
\@addtoreset{equation}{section}

\def\section{\@startsection{section}{1}{\z@}{3.5ex plus 1ex minus .2ex}
 {2.3ex plus .2ex}{\large\bf}}
\def\subsection{\@startsection{subsection}{2}{\z@}{2.3ex plus .2ex}
 {2.3ex plus .2ex}{\bf}}

\def\beq{\begin{equation} }
\def\eeq{\end{equation} }
\def\beqn{\begin{eqnarray} }
\def\eeqn{\end{eqnarray} }
\def\be{\begin{equation} }
\def\ee{\end{equation} }
\def\ben{\begin{eqnarray} }
\def\een{\end{eqnarray} }
\def\bear{\begin{eqnarray}}
\def\eear{\end{eqnarray}}

\def\Tr{{\rm Tr\,\,}}

\def\vev#1{\langle #1\rangle}

\def\MMSSM{ M_{\rm MSSM} }

\def\third{\frac{1}{3}}
\def\half{\frac{1}{2}}

\def\eps{\epsilon}
\def\UA{U(1)_{\rm A}}
\def\QA{Q^{({\rm A})}}
\def\mssm{SU(3)_C\times SU(2)_L\times U(1)_Y} 

\begin{document}

\begin{titlepage}
\samepage{
\setcounter{page}{1}
\rightline{CTP-TAMU-46/98}
\rightline{\tt hep-ph/9901203}
\rightline{January 1999}
\vfill
\begin{center}
 {\Large \bf  M--FLUENCES ON STRING MODEL BUILDING\\ }
\vfill
\vskip .4truecm
\vfill {\large Gerald B. Cleaver,\footnote{gcleaver@rainbow.physics.tamu.edu}}
\\
\vspace{.12in}
{\it        Center for Theoretical Physics,
            Dept.\  of Physics, Texas A\&M University,\\
            College Station, TX 77843\\}
\vspace{.06in}
{\it        and\\}
\vspace{.06in}
{\it        Astro Particle Physics Group,
            Houston Advanced Research Center (HARC),\\
            The Mitchell Campus, Woodlands, TX 77381\\}
\vspace{.025in}
\end{center}
\vfill
\begin{abstract}
Common to many classes of three family 
$SU(3)_C\times SU(2)_L\times U(1)_Y$ 
string models are an additional anomalous $\UA$ and numerous, often
fractionally charged, exotic particles beyond the minimal supersymmetric
standard model (MSSM). I present one string model with both of these 
properties. Using this model as an example, I show how sometimes 
(near) string scale vacuum expectation values that are both $D$--
and $F$--flat (to all orders in the superpotential)
can simultaneously break the anomalous $U(1)_A$ and give heavy mass
to all phenomenologically dangerous exotic states, decoupling them
from the low energy spectrum. For this model, some anomaly
cancelling directions actually decouple {\it all} MSSM exotics,
producing the MSSM as the effective field theory just below the string
scale. This model realizes Witten's conjecture of possible equivalence
between the string scale $M_{\rm S}$ and the MSSM unification scale $\MMSSM$.
\end{abstract}
\smallskip}
\begin{center}
{\it Talk presented at CPT '98, Bloomington, Indiana, 6--8 November 1998}
\end{center}
\end{titlepage}

\setcounter{footnote}{0}

%*****************************************************************************

\section{Anomalous $U(1)$ in Four--Dimensional String Models}

One of the underlying themes of this conference is possible 
generalizations to minimal supersymmetric standard model (MSSM) physics 
resulting from Lorentz and CPT violating effects. 
There have been very nice talks 
discussing how strings (or $M$--theory higher dimensional objects)
could produce such symmetry breaking effects. 
At this time I would like to focus on another type of symmetry breaking  
that,
as the six dimensions are compactified in string models, 
often accompanies the reduction  
of  ten--dimensional Lorentz symmetry 
to four--dimensional Lorentz symmetry. 
I am referring to the appearance of an anomalous local $\UA$ \cite{anoma}, 
for which there is a non--zero charge trace  
over the massless states in the effective low energy field theory,
\beq
         \Tr Q^{(A)} \neq 0\,\, .
\label{rotau1}
\eeq

In ten uncompactified dimensions, all local $U(1)$ symmetries 
in heterotic strings are 
embedded in either $E_8\times E_8$ or $SO(32)$ gauge groups and, therefore, 
are necessarily non--anomalous. However, producing the SM group
$\mssm$
from either of these larger groups via compactification
also frees several $U(1)_i$ (orthogonal to the SM group)
from their $E_8\times E_8$ or $SO(32)$ 
embeddings and allows them to become anomalous.  
Additionally, any of the six $U(1)_i$ corresponding
to the compactified dimensions may also become anomalous.
If more than one $U(1)_i$ is initially anomalous (which is the general case),
a unique rotation, 
\beq
\UA \equiv c_{\rm A}\sum_i \{\Tr Q^{(i)}\}U(1)_i\,\, ,
\label{rot}
\eeq
with $c_A$ a normalization coefficient,
places the entire anomaly into a single Abelian group, referred to here 
as $\UA$. 
All of the Abelian combinations orthogonal to $\UA$ become traceless. 
From hereon I will assume this rotation has been performed and will denote
the traceless combinations as $U(1)^{'}_{j}$. 

When an anomalous $\UA$ appears in a string model, 
there is a mechanism 
(aptly called the standard anomaly cancellation mechanism \cite{u1a}) 
whereby the $\UA$ is broken near the string scale. In the process,
a Fayet--Iliopoulos (FI) $D$--term,
\beq
      \eps\equiv \frac{g^2_s M_P^2}{192\pi^2}\Tr Q^{(A)}\, ,
\label{fidt}
\eeq
is generated in the effective Lagrangian, with
$g_{s}$ the string coupling and $M_P$ the reduced Planck mass, 
$M_P\equiv M_{Planck}/\sqrt{8 \pi}\approx 2.4\times 10^{18}$ GeV. 
The FI--term will break 
spacetime supersymmetry near the string scale unless a set of scalar
VEVs, $\{\vev{\varphi_m}\}$, of  fields $\varphi_m$
carrying anomalous charges $\QA_m$, can contribute a compensating
$D$--term,
$\vev{D_{A}(\varphi_m)} \equiv \sum_\alpha Q^{(A)}_m |\vev{\varphi_{m}}|^2$,
to cancel the FI--term, i.e.,
\beq
\vev{D_{A}}= \sum_m Q^{(A)}_m |\vev{\varphi_{m}}|^2 + \eps = 0\,\, .
\label{daf}
\eeq
%thereby restoring supersymmetry. 

A set of scalar VEVs satisfying eq.\ (\ref{daf})
is also constrained to maintain $D$--flatness for all non--anomalous Abelian
$U(1)^{'}_{j}$ symmetries as well,\footnote{I am only considering flat directions 
involving non--Abelian singlet fields solely. In cases where non--trivial 
non--Abelian representations are also allowed VEVs, generalized
non--Abelian $D$--flat constraints must also be imposed.}
\beq
\vev{D_{j}}= \sum_m Q^{(j)}_m |\vev{\varphi_{m}}|^2 = 0\,\, .
\label{dana}
\eeq
Each superfield $\Phi_{m}$ (containing the scalar field $\varphi_{m}$
and chiral superpartner) in the superpotential imposes further constraints 
on the set of scalar VEVs. 
$F$--flatness will be broken (thereby again destroying spacetime 
supersymmetry) at the scale of the VEVs unless,
\beq
\vev{F_{m}} = \vev{\frac{\partial W}{\partial \Phi_{m}}} = 0; 
\,\, \vev{W}=0\,\, .
\label{ff}
\eeq

Appearance of an anomalous $\UA$ in a string model can have profound
effects. 
An FI--term cancelling (and therefore supersymmetry restoring) flat direction 
of VEVs can drastically alter the 
phenomenology of a model in two ways.
First, the typical scalars taking on VEVs in a flat direction 
also carry charges of several non--anomalous $U(1)'_{j}$.
Through a generalized Higgs effect, the scalars
give near string--scale mass to the respective generators
of these gauge fields.
The standard anomaly cancellation mechanism thereby causes not just the
anomalous $\UA$ to be broken, but several non--anomalous $U(1)'_{j}$ as well.
Second, a non--renormalizable superpotential term,  
$\frac{\lambda}{M_{\rm S}^{n-3}} \Phi_{1}\Phi_{2}\Phi_{3}\dots \Phi_{n}$,
formed from $n>3$ superfields 
produces a new effective mass term, 
$\frac{\lambda}{M_{\rm S}^{n-3}}
\langle \Phi_{1}\Phi_{2}\Phi_{3}\dots \Phi_{n-2}\rangle \Phi_{n-1} \Phi_{n}$,
if $n-2$ of the fields take on VEVs, 
or produces a new effective Yukawa term,
$\frac{\lambda}{M_{\rm S}^{n-3}}
\vev{\Phi_{1}\Phi_{2}\Phi_{3}\dots \Phi_{n-3}}\Phi_{n-2}\Phi_{n-1}\Phi_{n}$,
if $n-3$ take on VEVs. ($\lambda$ is a generic non--renormalizable 
coupling coefficient.)

\section{Three Generation 
$SU(3)_C\times SU(2)_L\times U(1)_Y\times \prod_i U(1)_i$ String Models}

As mentioned,
in four dimensions, a quasi--realistic three generation $\mssm$ 
heterotic string model has several additional Abelian group factors,
$U(1)_i$, along with a hidden sector non--Abelian gauge group 
$G_{\rm hid}$.  
When this type of string model is compactified
from ten to four dimensions
via a Calabi--Yau (CY) manifold or $N=2$ minimal model,
all of the extra $U(1)_i$ remain non--anomalous. 
On the other hand, when this compactification occurs through
bosonic lattices, orbifolds, or free fermions,  
an anomalous $\UA$ generically appears.
Thus, while CY and $N=2$ compactified models can be thought of as 
``What you see is what you get'' models,
lattice, orbifold, and free fermion models cannot be regarded as such.
For the latter classes of models, the massless spectrum and
gauge groups before anomaly cancellation may be very different from
the corresponding ones after a flat direction set of VEVs is chosen
(non--perturbatively).

   These dynamical, FI--term induced string model transformations 
may, in fact, be a very valuable tool for producing 
phenomenologically viable string models. The reason for this relates to 
a problematic aspect of generic three generation $\mssm$
string models. 
Exotic MSSM states, many carrying fractional electric charge, 
seem an ubiquitous feature of such models \cite{fc,aft}.   
Most of these exotics, if they remain massless down to the electroweak scale,
signify unphysical phenomenology, thereby disallowing a model containing 
them. 
Enhancing the probability of this occurring are the 
``string--selection rules.'' 
These are additional constraints on 
superpotential terms beyond standard gauge invariance.   
String selection rules often forbid superpotential terms, 
otherwise allowed by gauge invariance, that could 
generate large mass for an exotic via couplings with 
flat direction VEVs \cite{cceelw}.
This generally makes decoupling of all 
dangerous exotic fields from the low energy effective field theory 
difficult. 

\section{A String Derived MSSM}

Recently a free fermionic string model constructed several years ago 
\cite{fny,fc} was found to contain certain flat  
(that is, $D$-- and $F$--flat to all finite orders in the 
superpotential \cite{cceel})
directions of near string scale magnitude  
that {\it simultaneously} cancel the FI $D$--term and 
give (near) string scale mass to {\it all} exotics (including
those with fractional electric charge) \cite{cfn1}. 
Prior to VEVs being turned on, 
the gauge group of this model is
$\mssm \times U(1)_A \times \prod_{j=1}^{10} U(1)^{'}_{j} \times 
SU(3)_H\times SU(2)_H\times SU(2)'_H$.
Under any of these special flat directions, exactly three
$U(1)^{'}_{j}$ survive. The gauge group thus reduces to
$\mssm \times \prod_{j'=1}^{3} U(1)^{'}_{j'} \times 
SU(3)_H\times SU(2)_H\times SU(2)'_H$.

In the pre--VEV stage, the model contains several massless MSSM exotics:
one $SU(3)_C$ vector--like pair of triplets with electric charges 
$Q_{elec}= \pm \third$;
ten $SU(2)_L$ doublets, four of which carry $Q_{elec}= \pm\half$;
and 16 $SU(3)_C\times SU(2)_L$ singlets, eight with
$Q_{elec}= +\half$ and eight with $Q_{elec}= -\half$;
one vector--like pair of hidden sector $SU(2)_H$ doublets with
$Q_{elec}= -\half$, and a similar pair of $SU(2)'_H$ doublets.
The specific flat directions generate mass
through unsuppressed renormalizable superpotential terms 
for all of the exotics, except for the $SU(3)_C$ vector--like triplet pair.
The $SU(3)$ triplet pair receives mass through a suppressed 
fifth order term and, thus, its mass scale should be suppressed a bit
below that of the other states.
The FI--term induced mass scale of the fields was found to be of the 
order $\sim 7\times 10^{16}$ GeV.

One physical characteristic distinguishing between these various flat
directions is the set of additional non--Abelian singlets and 
hidden sector non--Abelian non--singlets that take on (near) string--scale 
mass. Typically, slightly less than half of the remaining 47 singlets become 
massive.
For the example flat direction discussed in ref.\ \cite{cfn1}, the number 
is 19, with
the corresponding mass terms being unsuppressed for 15 of these. 
Four singlets receive (suppressed) mass terms at fifth order.
Slightly more than half of the hidden sector non--singlets also
receive induced masses. For these non-singlets, 
more masses are generated through higher order terms than through
the renormalizable terms.
When the flat direction of ref.\ \cite{cfn1} is applied,       
18 of the remaining 30 hidden sector non--singlet states become
near string scale massive. Eight of these 18 states gain mass through
renormalizable terms, six from fourth order terms, and 
four via fifth order terms. 

\section{String and MSSM Scale Unification}

Assuming the spectrum of the MSSM above the electroweak scale,
unification of the $SU(3)_C$, $SU(2)_L$ and $U(1)_Y$ running 
couplings occurs at a scale  
$M_{\rm U} = \MMSSM\approx 2.5 \times 10^{16}$ GeV. However, 
for several years \cite{kap} the general perturbative string 
prediction of the scale at which all gauge couplings merge
has been on the order of $M_{\rm S}\approx 5\times 10^{17}$ GeV.
Several perturbative solutions have been proposed \cite{krdrev} 
to resolve the apparent factor of 20
inequity between the two scales, $\MMSSM$ and $M_{\rm S}$.
Typically these proposals attempt to 
(i) raise the $SU(3)_C$, $SU(2)_L$ and $U(1)_Y$ unification scale 
$M_{\rm U}$ above the MSSM scale $\MMSSM$ 
through the effects of intermediate scale MSSM exotics on the
running couplings, 
(ii) lower the string scale $M_{\rm S}$ to the MSSM scale $\MMSSM$
via threshold effects 
from the infinite tower of massive string states, 
(iii) run a unified MSSM coupling to the string scale via   
a grand unification theory, or 
(iv) various combinations of (i) through (iii). 
However, Witten has recently suggested an $M$--theory mechanism 
that offers a non--perturbative resolution to the apparent 
scale misalignment \cite{mth}.  
This conjecture maintains the successful MSSM 
prediction and equates the string scale to the MSSM scale,
$M_{\rm S}= \MMSSM\approx 2.5 \times 10^{16}$ GeV.
Thus, this conjecture suggests that the observable gauge group 
just below the string scale should be 
$SU(3)_C\times SU(2)_L\times U(1)_Y$ and the spectrum of the 
observable sector should consist solely of the MSSM spectrum.
The model of \cite{fny,fc,cfn1} appears to be the first realization
of an actual string--derived MSSM.

Detailed analysis of this model is underway and will be presented 
in forthcoming papers \cite{cfn2}. 
In particular, for each MSSM--generating flat direction,  
we will examine the variations in
(i) the textures of the MSSM mass matrices,
(ii) the non--Abelian singlet and hidden sector non--singlet 
low energy spectrums, and
(iii) the hidden sector effective Yukawa and non--renormalizable terms.

\section{Acknowledgements}

G.C.\  thanks the coordinator of CPT '98, Alan Kosteleck\' y,
and his staff for organizing a very enjoyable and educational conference.
G.C. also thanks his collaborators Alon Faraggi and Dimitri Nanopoulos
for valuable discussions.
This work is supported in part by DOE Grant No.\ DE--FG--0395ER40917.

%*********************************************************************
\def\NPB#1#2#3{{\it Nucl.\ Phys.}\/ {\bf B#1} (#2) #3}
\def\NPBPS#1#2#3{{\it Nucl.\ Phys.}\/ {{\bf B} (Proc.\ Suppl.) {\bf #1}}
 (#2) #3}
\def\PLB#1#2#3{{\it Phys.\ Lett.}\/ {\bf B#1} (#2) #3}
\def\PRD#1#2#3{{\it Phys.\ Rev.}\/ {\bf D#1} (#2) #3}
\def\PRL#1#2#3{{\it Phys.\ Rev.\ Lett.}\/ {\bf #1} (#2) #3}
\def\PRT#1#2#3{{\it Phys.\ Rep.}\/ {\bf#1} (#2) #3}
\def\MODA#1#2#3{{\it Mod.\ Phys.\ Lett.}\/ {\bf A#1} (#2) #3}
\def\IJMP#1#2#3{{\it Int.\ J.\ Mod.\ Phys.}\/ {\bf A#1} (#2) #3}
\def\nuvc#1#2#3{{\it Nuovo Cimento}\/ {\bf #1A} (#2) #3}
\def\RPP#1#2#3{{\it Rept.\ Prog.\ Phys.}\/ {\bf #1} (#2) #3}
\def\etal{{\it et.\ al\/}}
%*********************************************************************


\begin{thebibliography}{99}

\bibitem{anoma} For general discussions of anomalous $U(1)$ in string models
see, e.g.,\\ 
T. Kobayashi and H. Nakano, \NPB{496}{1997}{103}, [hep-th/9612066];\\
G.B.~Cleaver, \NPBPS{62A-C}{1998}{161} [hep-th/9708023];\\
G.B.~Cleaver and A.E.~Faraggi, UFIFT-HEP-97-28, UPR-0773-T, [hep-ph/9711339];\\
A.E.~Faraggi, \PLB{426}{1998}{315}, [hep-ph/9801409];\\
L.E.~Ib\'{a}\~{n}ez, R.~Rabadan, and A.M.~Uranga, [hep-th/9808139];\\  
P.~Ramond, Proceedings of Orbis Scientiae '97 II, 
Dec.~1997, Miami Beach, [hep-ph/9808488];\\
W.~Pokorski and G.G.~Ross, [hep-ph/9809537];
and references within.

\bibitem{u1a}{M. Dine, N. Seiberg and E. Witten, \NPB{289}{1986}{585};\\
J. Atick, L. Dixon and A. Sen, \NPB{292}{1987}{109}.}

\bibitem{fc}{A.E. Faraggi, \PRD{46}{1992}{3204}.}

\bibitem{aft}{A.E. Faraggi, \NPB{387}{1992}{239}, [hep-th/9208024];
Proceedings of PASCOS 92, [hep-ph/9301220];
%``Toward the Classification of Realistic Free Fermionic Models,'' 
[hep-th/9708112];
Proceedings of Strings 96, [hep-ph/9608420];\\
J. Elwood and A. Faraggi, \NPB{512}{1998}{42}, 
[hep-ph/9704363].}

\bibitem{cceelw}{G. Cleaver, M. Cveti\v c, J.R. Espinosa, L. Everett, 
  P. Langacker and J. Wang, \PRD{59}{1999}{055005}, [hep-ph/9807479];  
  \PRD{59}{1999}{115002}, [hep-ph/9811355].}

\bibitem{cceel}{G. Cleaver, M. Cveti\v c, J.R. Espinosa, L. Everett, and
  P. Langacker, 
\NPB{525}{1998}{3}, [hep-th/9711178];
``Flat Directions in Three Generation String Models'', 
CERN-TH/98-154, UPR-0784-T, IEM-FT-173/98, [hep-ph/9805133].}

\bibitem{fny}{A.E. Faraggi, D.V. Nanopoulos, and K. Yuan, \NPB{335}{1990}{347}.}

\bibitem{cfn1}{G.B. Cleaver, A.E. Faraggi, and D.V. Nanopoulos,
``String Derived MSSM and M--Theory Unification,''
ACT-11/98, CTP-TAMU-45/98, TPI-MINN-98/24, UMN-TH-1729-98,
[hep-ph/9811427].}

\bibitem{kap}{V. Kaplunovsky, \NPB{307}{1988}{145}; 
Erratum {\bf B382} (1992) 436.;\\
V. Kaplunovsky and J. Louis, \NPB{444}{1995}{191}.}

\bibitem{krdrev}{General references for and reviews of the proposals, 
including discussions of their varying levels of success, can be found in:\\ 
K. Dienes and A. Faraggi, \NPB{457}{1995}{409}, [hep-th/9505046];\\
K. Dienes, \PRT{287}{1997}{447}, [hep-th/9602045].}

\bibitem{mth}{E. Witten, \NPB{471}{1996}{135}; For a review see,\\
D.V.  Nanopoulos, ``M--Phenomenology,'' CTP-TAMU-42/97, ACT-15/97,
[hep-th/9711080].}

\bibitem{cfn2}{G.B. Cleaver, A.E. Faraggi and D.V. Nanopoulos,
``A Minimal Superstring Standard Model I: Flat Directions,''
ACT-2/99, CTP-TAMU-12/99, TPI-MINN-99/22, UMN-TH-1760-99,
[hep-ph/9904301]; 
``A Minimal Superstring Standard Model II: The Phenomenology,''
paper in preparation.} 

\end{thebibliography}
\end{document}